\documentclass{aa}

\usepackage{txfonts}
\usepackage{graphicx}
\usepackage{rotating}
\usepackage{natbib}
\usepackage{lscape}
\usepackage{multirow}

\bibpunct{(}{)}{;}{a}{}{,} %to follow the A&A style

\newcommand{\gapprox}{$\stackrel {>}{_{\sim}}$}   %greater/less approx.
\newcommand{\lapprox}{$\stackrel {<}{_{\sim}}$}

\begin{document}

\title{Investigating the past history of EXors: the cases of V1118 Ori, V1143 Ori, and NY Ori}

\author{R. Jurdana-\v{S}epi\'{c}\inst{1},
        U. Munari\inst{2},
        S. Antoniucci\inst{3},
        T. Giannini\inst{3},
        G. Li Causi\inst{3,4},
        D. Lorenzetti\inst{3}
                }

\institute{ Physics Department, University of Rijeka, Radmile Matej\v{c}i\'{c}, 51000 Rijeka, Croatia \and
            INAF - Osservatorio Astronomico di Padova - via dell’Osservatorio 8, Asiago(VI) 36012, Italy \and
            INAF - Osservatorio Astronomico di Roma - Via Frascati, 33 - Monte Porzio Catone 00078, Italy \and
            INAF - Istituto di Astrofisica e Planetologia Spaziali - Via Fosso del Cavaliere, 100 - Roma 00133, Italy
                        } 

\offprints{Dario Lorenzetti, \email{dario.lorenzetti@oa-roma.inaf.it}}
\date{Received date / Accepted date}
\titlerunning{Past history of EXors}
\authorrunning{Jurdana-\v{S}epi\'{c} et al.}

\abstract
{EXor objects are young variables that show episodic variations of brightness commonly associated to enhanced accretion outbursts.}
{With the aim of investigating the long-term photometric behaviour of a few EXor sources, we present here 
data from the archival plates of the Asiago Observatory, showing
the Orion field where the three EXors V1118, V1143, and NY are located.}
{A total of 484 plates were investigated, providing a total of more than 1000 magnitudes for the three stars, which cover 
a period of about 35 yrs between 1959 to 1993. We then compared our data with literature data.} 
{Apart from a newly discovered flare-up of V1118, we identify the 
same outbursts already known, but we provide two added values: ({\it i}) a long-term sampling 
of the quiescence phase; and ({\it ii}) repeated multi-colour observations (BVRI bands).
The former allows us to give a reliable characterisation of the quiescence, which represents
a unique reference for studies that will analyze future outbursts and the physical changes induced by these events.
The latter is useful for confirming whether the intermittent increases of brightness are accretion-driven 
(as in the case of V1118), or extinction-driven (as in the case of V1143). Accordingly, doubts arise about 
the V1143 classification as a pure EXor object. Finally, although our plates do not
separate NY Ori and the star very close to it, they indicate that this EXor did not undergo any major outbursts during our 40 yrs of monitoring.}
{}


\maketitle

\section{Introduction}

Among the low-mass protostars, the EXor-type objects (hereafter EXor -  Herbig 1989) are considered a peculiar sub-class of the classical T Tauri stars (CTTS). These sources undergo episodic outbursts of brightness (typically of three to five magnitudes at optical wavelengths) caused by sudden variations of the mass accretion rate that obeys the magnetospheric accretion rules (Shu et al. 1994). At the moment, the known EXors are only few tens (Audard et al. 2014) usually discovered by chance. Indeed, all the long term photometric surveys carried
out so far on CTTS samples confirm that they present, on average, small amplitude variations (typically \lapprox 1 mag). These results tend to support
a certain peculiarity of the EXors, but to what extent their accretion events are unfrequent manifestations of a common phenomenology (Lorenzetti et al 2012) is not firmly ascertained. \

EXors present short outburst (months--one year) with a recurrence time of years, associated with accretion rates of the order of 10$^{-6}$-10$^{-7}$ M$_{\odot}$~yr$^{-1}$, and characterised by emission-line spectra (e.g. Herbig 2008; Lorenzetti et al. 2009; K\'{o}sp\'{a}l et al. 2011; Sicilia-Aguilar et al. 2012; Antoniucci et al. 2013, Antoniucci et al. 2014). Such features, and in particular the burst cadence, make EXors the ideal candidates for a long-term monitoring, which should be effective in comparing the properties of several subsequent events and evidence their similarities or differences. Following this observational approach, compelling constraints can be provided for the models currently available. Indeed, the EXors phenomenology has been so far interpreted by borrowing the theoretical approaches developed to study the FUor events (Hartmann \& Kenyon 1985), but a detailed model of the disk structure and its evolution does not exist yet for EXors stars. D'Angelo \& Spruit (2010) provided quantitative predictions for the episodic accretion of piled-up material at the inner edge of the disk that, however, are largely
incompatible with the observations. Hence, the mechanism responsible for the onset of EXor accretion outbursts remains unknown to date. Two proposed scenarios involve essentially either disk instabilities (gravitational - e.g. Adams \& Lin 1993, or thermal - e.g. Bell $\&$ Lin 1994) or perturbation by an external body (a close encounter in a binary system - e.g. Bonnell \& Bastien 1992, or presence of a massive planet - e.g. Lodato $\&$ Clarke 2004).\

The young star V1118 Ori, one of the classical EXor objects, represents a very suitable target for 
such a kind of long term monitoring. During its recent history V1118 Ori underwent six documented outbursts, each lasting a couple of years (1982-84, 1988-90, 1992-94, 1997-98, 2004-06, 2015-on going). Account for the first five events is given in Parsamian et al. (1993), Garc\'{i}a Garc\'{i}a \& Parsamian (2000), Herbig (2008) and references therein, Audard et al. (2005, 2010), and Lorenzetti et al. (2006, 2007). The properties shown during the long quiescence period before the last eruption, firstly detected around Sept. 2015, are described in Lorenzetti et al. (2015), Giannini et al. (2016), and
Giannini et al. (in preparation). In this latter paper, a comparison of all the recent outbursts is presented, with the aim of investigating if any periodicity is recognizable among the recurrent events. However, so far only a couple of outbursts have been sampled at an accuracy level that allows us to have some indication on this aspect. A firm ascertainment of the existence of a periodicity could speak in favour of a disk
instability generated by an external body as the most likely mechanism for the mass accretion variations.
To this end, we have investigated the optical plates collected by the Asiago (Italy) Schmidt telescopes during a timespan of about 40 years (1958-1998), which are
able to provide the most accurate long-term light curve of V1118 Ori available before the advent of both CCD and near-IR arrays.\

Beside V1118 Ori, two other EXors, namely V1143 Ori and NY Ori, are also located in the Orion Nebula Cluster (ONC), thus appearing in the same archival plates of the Asiago observatory. By analysing these large number of plates we have therefore the great advantage of investigating the past history of two additional systems with the same method and accuracy. 
V1143 is a very active variable (see Herbig 2008, and references therein) and presents ranges of photometric fluctuations of $\sim$19-14.3 mag and 16.7-13.5 mag in B/pg and V/vs bands, respectively. 
NY Ori (= Parenago 2119) is a young star located 5.5 arcsec SE of the much optically brighter source V566 Ori (= Parenago 2118). Its photometric history and variability is documented (Herbig 2008) by a series of plates collected during short periods (1905-19; 1948-49; 1973) with very sparse and occasional sampling, which provide a range of variability in the visual band between 13.3 and 16 mag. 
A previous plate analysis of V1118 Ori and V1143 Ori was presented by Paul et al. (1995), but the extremely low number of resulting photometric data points (less than ten in 40 years), hampers a proper time coverage. 

In any case, the relevance of the plates analysis for studying the eruptive variables is irrefutable, as recently demonstrated for the EXor GM Cep (Xiao et al. 2010) and the FUor V960 Mon (Jurdana-\v{S}epi\'{c} \& Munari, 2016).\
Noticeably, the chance of studying the historical light curves of EXors located in the Orion nebula is offered uniquely by the Asiago plates. Indeed, other remarkable archives (e.g. Harvard and Sonneberg) present some difficulties in this respect: 
firstly, they are not as much deep to investigate the quiescence phase; 
secondly, they cover almost exclusively the B or blue-unfiltered bands;
thirdly, they have been obtained with short focal length astrographs that typically squeeze the plate scale making
indistinguishable the stars inside the nebula. 
To our best knowledge, plate-based monitoring of other EXors as faint as V1118 Ori is not available in literature. 
The features mentioned above suggest that the Asiago archive might be a reference source
for investigating the past history of both recently discovered EXors and those that will be discovered in the next future. 

The present paper is organized as follows: the adopted method and the obtained BVRI photometry are presented in Sections 2 and 3.
Section 4 gives the analysis and discussion of the obtained results, while our concluding remarks are given in Sect. 5.

\section{Data acquisition} 

\subsection{Archive plates}

Two Schmidt telescopes are operated at the Asiago observatory.  The smaller
one (SP: 40/50 cm, 100 cm focal length) collected 20,417 plates between 1958 to
1992 (covering a circular area 5$^\circ$ in diameter), and the larger one
(SG: 67/92 cm, 208 cm focal length) 18,811 plates from 1965 to 1998 (imaging a
5$^\circ\times$5$^\circ$ portion of the sky).  The Asiago Schmidt plate
collection thus span 40 years, covering in particular the {\it Menzel Gap}
during which acquisition of photographic plates was temporarily halted at
Harvard Observatory.  Since the 1990s, large format CCD cameras have replaced
photographic plates as detectors, with the plates preserved in controlled
conditions. 

The Asiago Schmidt plates have not been exposed over the whole sky but
instead on selected targets (in particular star forming regions, clusters of
galaxies, galaxies of the Local Group), for which therefore exist a long
record of observations.  The Orion nebula - around which our targets are
located - has been a favoured hit.  Nearly all plates from both telescopes
go very deep, $B$$\sim$18.5 and $B$$\sim$17.8 mag being the typical limiting
magnitude for blue sensitive plates exposed with the SG and SP
telescope, respectively.

The majority of the plates were exposed as 103a-O + GG13, 103a-E + RG1 and IN +
RG5 combinations of Kodak plates and Schott astronomical filters, matching
the prescription for the Johnson-Cousins $B$, $R_{\rm C}$ and $I_{\rm C}$
photometric bands (Moro\& Munari 2000), respectively.  A large number of
plates were exposed as unfiltered 103a-O, thus covering both the Johnson $B$
and $U$ bands thanks to the high ultraviolet transparency of the UBK-7
corrector plates at both Schmidt telescopes.  For low temperature and/or
reddened objects, especially if they were observed at large airmass, the amount of
proper $U$-band photons collected by an unfiltered 103a-O plate is however
minimal compared to those arriving through the $B$-band portion of the
interval of sensitivity of 103a-O emulsion.  In such conditions
(and provided that the selected comparison stars are themselves of low
temperature and/or high reddening), 103a-O + GG13 pairs and unfiltered
103a-O plates are almost equally well replicating the standard Johnson $B$
band (Munari \& Dallaporta 2014).

The plates in the proper Johnson $V$ band obtained as
103 a-D + GG14 combination, are low in number, and equally infrequent are unfiltered
panchromatic plates as 103 a-G, 103a-D, Tri-X and Pan-Roy.  In the present
paper, these unfiltered panchromatic plates have been treated as $V$-band
plates (even if they lack the coupling to a high-pass filter such as GG11 or
GG14) because of the low temperature and reddening of our targets and
generally the high airmass ($\sim$1.8) of the corresponding observations
which prevent $U-$ and $B-$band photons to be collected in any significant
number by unfiltered panchromatic plates.

\begin{figure*}
\centering
\includegraphics[width=15cm, trim = 0cm .5cm 0cm 1cm, clip]{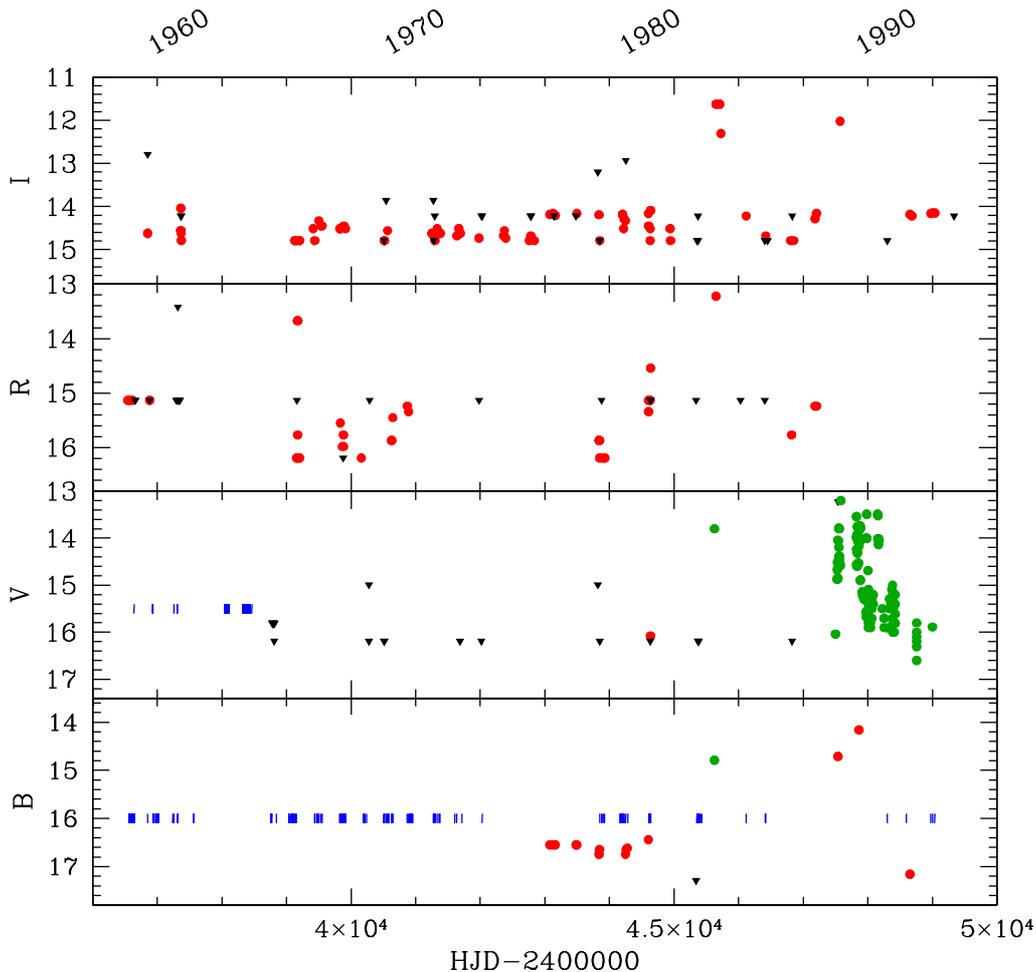}
\caption{\label{V1118_light:fig} $BVRI$ light curves of V1118 Ori. Upper limits are given as solid black triangles. In the two bottom panels ($B$ and $V$), blue vertical bars indicate the dates when the nebular background around V1118 Ori was too bright (see text Sect.~2.2). Green points refer to literature measurements reported by Parsamian et al. (1993) and Garc\'{i}a Garc\'{i}a \& Parsamian (2000).}
\end{figure*}

\begin{figure*}
\centering
\includegraphics[width=15cm, trim = 0cm .5cm 0cm 1cm, clip]{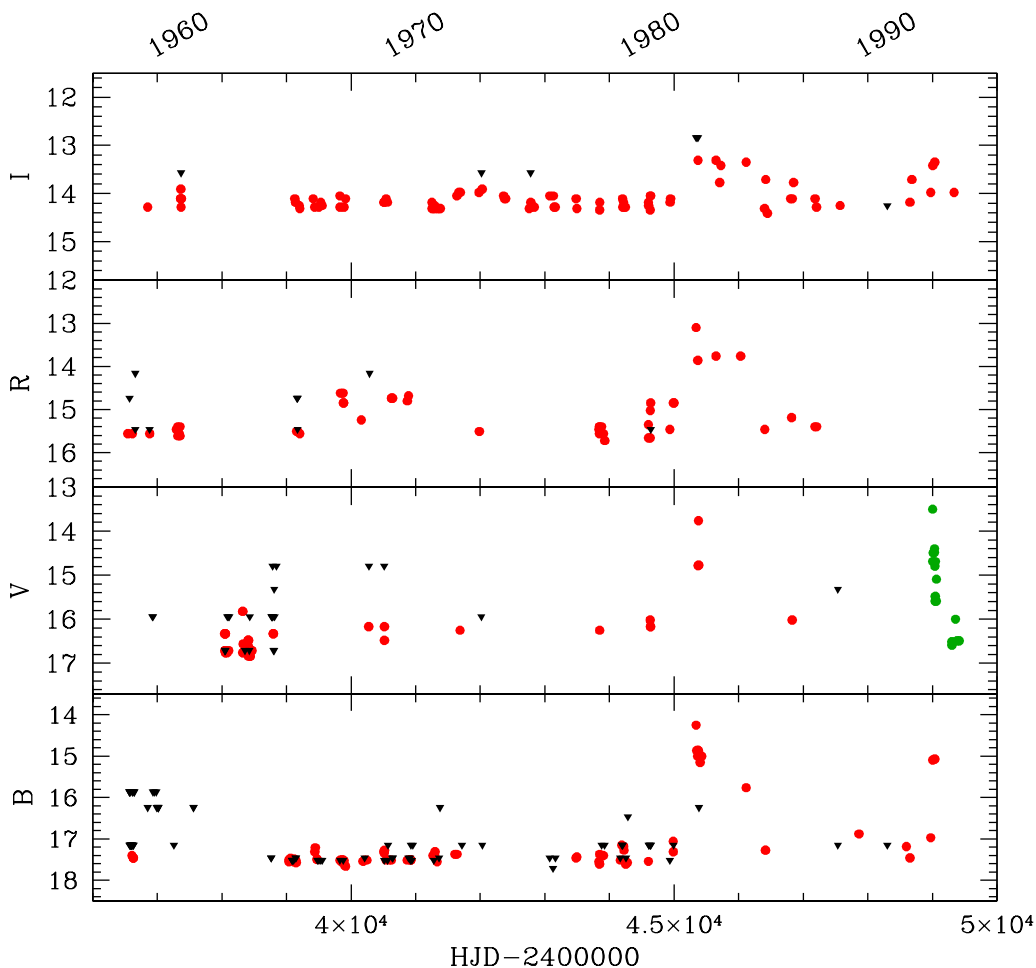}
\caption{\label{V1143_light:fig} $BVRI$ light curves of V1143 Ori. Upper limits are given as solid black triangles. Green points are measurements taken from by Mampaso \& Parsamian (1995).}
\end{figure*}

\begin{figure*}
\centering
\includegraphics[width=15cm, trim = 0cm .5cm 0cm 1cm, clip]{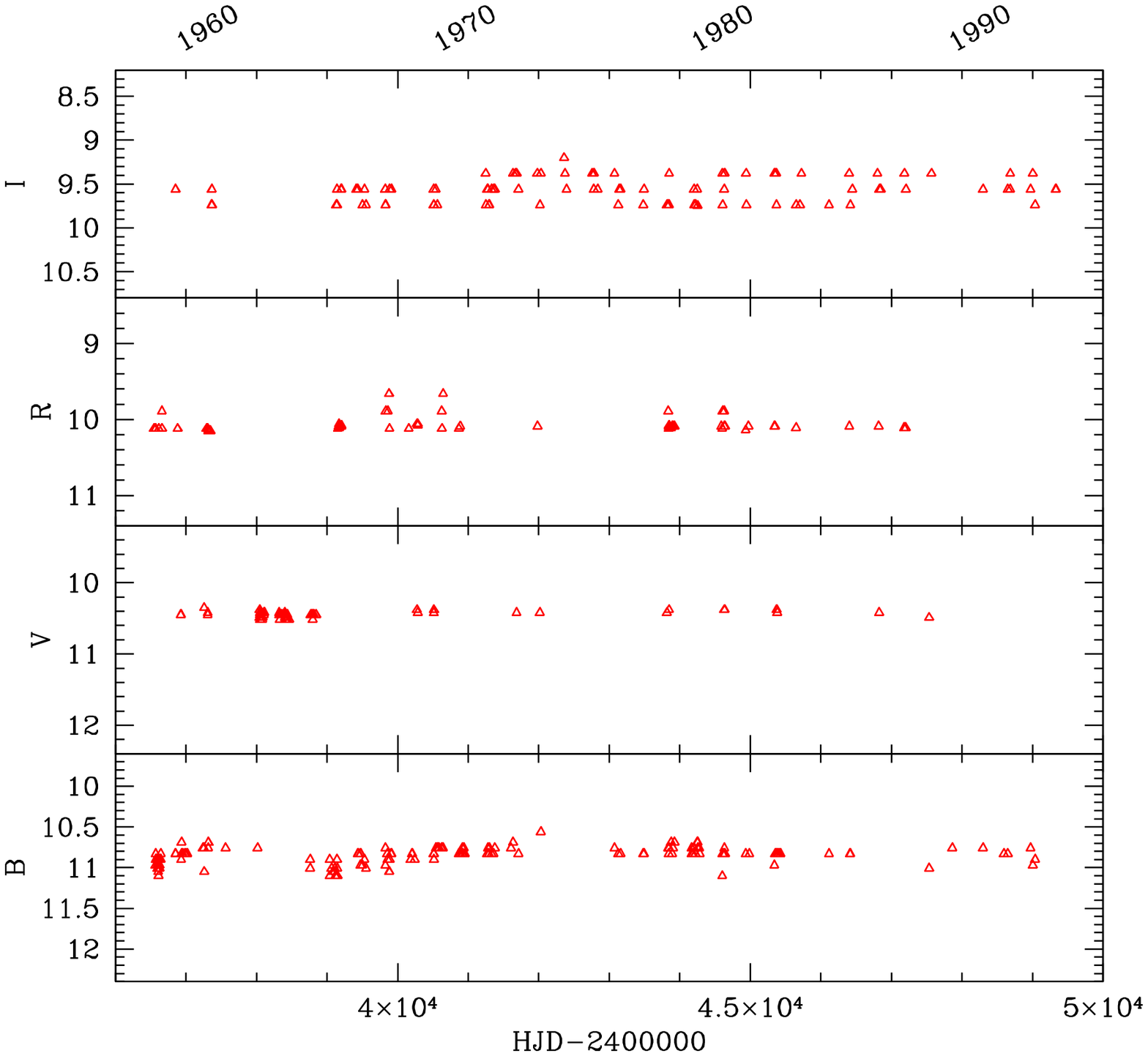}
\caption{\label{NY_light:fig} $BVRI$ light curves of NY Ori.}
\end{figure*}

\subsection{Brightness measurement}

To derive the brightness of our targets, we compared them at a high quality
Zeiss microscope against a local photometric sequence established around
each target.  Such a sequence, composed of stars of roughly the same colour
as the variable and widely distributed in magnitude so to cover both
quiescence and outburst states, was extracted primarily from the APASS
$B$$V$$g'$$r'$$i'$ all-sky survey (Henden et al.  2012, Henden \& Munari
2014), with porting to Landolt $R_{\rm C}$ and $I_{\rm C}$ bands following
Munari (2012) and Munari et al. (2014) prescriptions.  To evaluate the
measurement errors a number of plates were re-measured after several days
when all memories had vanished from the observer, and independently by
different observers.  The typical error is 0.1 mag, comparable to that
intrinsic to the photographic plate itself so that the observer adds little
to it.  When a larger error is reported, that is usually associated to
presence of a fogged background, a trailed guiding, or poor seeing.

A total of 484 plates imaging the Orion nebula were retrieved from the
Asiago Schmidt plate archive.  After inspection, 44 of them were rejected
for various reasons (too short exposures, unsuitable emulsion or filter
combinations, damages during development or storing, etc.), leaving a total
of 440 plates ready for measurement, spanning the time interval from Dec 7,
1958 to Dec 12, 1993.  The results are given in Table~\ref{photometry:tab} where the
columns provide date/UT/HJD of observation, photographic
emulsions and filters, plate number and telescope, the estimated magnitude or plate limit, and the
corresponding errors. When the variable is too faint to be detected, the
limiting magnitude is listed as that of the faintest of the stars in the
comparison sequence which is clearly visible.

The faint variables V1118 Ori and V1143 Ori are superimposed to the Orion
nebula, which greatly disturbs their detection and measurement.  All
emission from the nebula come from few extremely intense emission lines,
mainly [OIII] 4959, 5007 \AA\ and the [NII]+H$\alpha$ complex.  A minimal
variance in the transmission of the emulsion+filter combination at these
wavelengths causes large changes in the brightness background.  For this
reason, in a significant number of plates the nebular background around
V1118 Ori is so bright that it is impossible to estimate the brightness in
quiescence of such a faint variable star. For these plates (identified in Table~\ref{photometry:tab} 
and whose date of observation is depicted in Figure~\ref{V1118_light:fig})
it was only possible to exclude that the target was at that time going through a bright outburst.  
NY Ori is a member of a close optical pair with brighter 2MASS
J05353579-0512205 (=V566 Ori) at 5 arcsec distance.  On the Palomar Schmidt
plates they can be percept as a heavily blended pair, with the two stars not
measurable separately, while at the shorter focal length of the Asiago
Schmidt telescopes the pair is merged into the image of a unresolved single
star.
Therefore the measurements of NY Ori reported in Table~\ref{photometry:tab} refer to
the combined pair of stars.

\section{Historical light curves} 

The $BVRI$ light curves of V1118, V1143, and NY Ori corresponding to their plate photometry given in Table~\ref{photometry:tab}, are given in Figures~\ref{V1118_light:fig}, \ref{V1143_light:fig}, and \ref{NY_light:fig}, respectively. In Table~\ref{ranges:tab} some statistics are 
provided for the three sources in each band, namely the number of observations, the median value, (basically the quiescence magnitude), 
together with the standard deviation of the data points, and the peak brightness.
Notably, we provide one of the best sampled data set ever obtained of the quiescent phase of the investigated sources. Indeed, albeit observations are concentrated within the seasonal periods of observability, their number is significantly high (between 50 and 100) in many cases, allowing us to infer meaningful averaged values for the quiescence level. 
Determining this level is fundamental to have a solid reference for %accurately computing physical changes, once the outburst values are obtained. 
future observations that will analyse new outbursts and investigate the physical changes induced on the system by these enhanced accretion events.
Finally, we note how these sources (especially V1118 and V1143) present, in quiescence, a level of modest variability. This may be quantified by considering the standard deviation of the measurements (0.2-0.4 mag), which is comparable with that of the classical T Tauri (CTTS).

\begin{figure*}
\centering
%\vspace*{-7cm}
\includegraphics[width=15cm, trim = 0cm 0cm 0cm 9cm, clip]{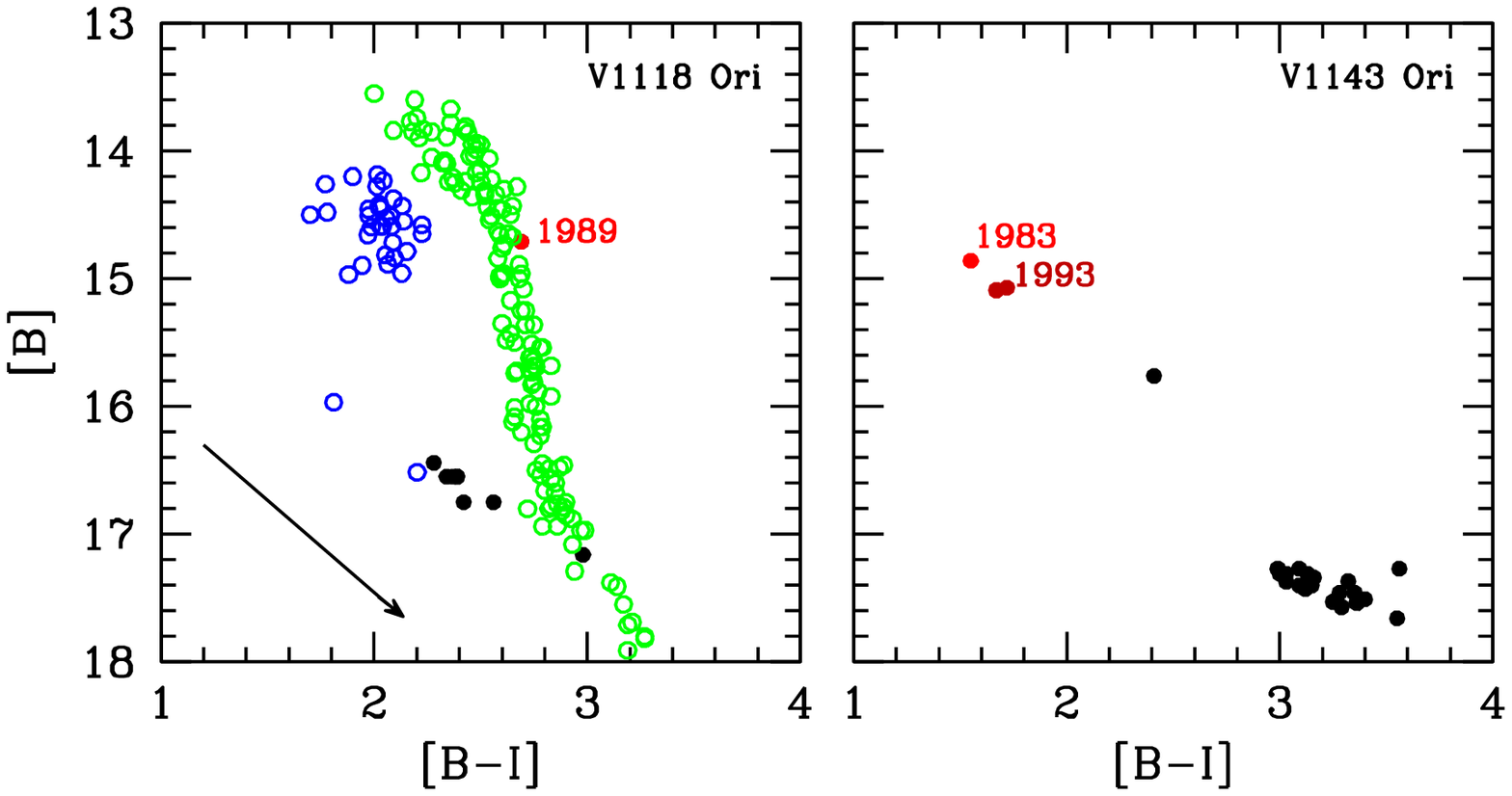}
\caption{\label{B_BI_plot:fig} $B$ vs [$B-I$] colour-magnitude plot of V1118 (left) and V1143 Ori (right). In the left panel open circles (in blue) refer
to our data still unpublished (Giannini et al. in preparation). Solid circles (black and red) refer to the present data. Green circles are given for comparison purposes and refer to a recent outburst of V1118 monitored by Audard et al. (2005). In the lower-left corner the arrow indicates an
extinction of A$_V$ = 1 mag, according to the law by Rieke \& Lebofsky (1985).}
\end{figure*}

%%%%%%%%%%%%%%   TABLE 1 - PHOTOMETRY %%%%%%%%%%%%%%%%%%%%%%%%%%%%%%%%%
%\setcounter{table}{0}
 
\begin{table*}
\caption{$BVRI$ plate photometry of NY, V1143, V1118 in Orion. Columns provide: date and central UT of any exposure, plate emulsion, the adopted filter, the telescope, the plate number, and the magnitude derived for the three sources with the estimated error (see text for further details). For V1118 Ori, the asterisks indicate plates for which it was only possible to rule out that the source was going through a bright outburst.
\label{photometry:tab}} 
\begin{tiny}
\begin{tabular}{l|cc|cccc|cc|cc|cc|}
\hline
\hline
  &   &   &   &   &  &  & \multicolumn{2}{c|}{NY Ori} & \multicolumn{2}{c|}{V1143 Ori} & \multicolumn{2}{c}{V1118 Ori} \\
HJD  &  date  &  UT  &  emulsion & filter  & tel & plate & mag & error & mag & error & mag & error \\
%                                                                                                          NY Ori                    V1143 Ori                V1118 Ori (buone)           
%  HJD              date             UT          emuls         filt       tel       plate              mag      err                mag       err               mag       err       
\hline                                                                                                                     
2436545.43698  &    1958-12-07  &    22:22  &    103 a-E  &    RG 1  &     SP   &     59   &         $R$  = 10.12 &   0.10   &       $R$  = 15.56  &  0.10    &      $R$  = 15.13  &  0.20   \\ 
2436567.31360  &    1958-12-29  &    19:25  &    103 a-O  &    ...   &     SP   &     66   &         $B$  = 10.97 &   0.10   &       $B$ $>$15.86  &  0.10    &      $*$           &         \\ 
2436570.29404  &    1959-01-01  &    18:57  &    103 a-O  &    ...   &     SP   &     70   &         $B$  = 10.83 &   0.10   &       $B$ $>$15.86  &  0.10    &      $*$           &         \\ 
2436570.36626  &    1959-01-01  &    20:41  &    103 a-E  &    RG 1  &     SP   &     73   &         $R$  = 10.12 &   0.10   &       $R$ $>$14.74  &  0.10    &      $R$  = 15.13  &  0.20   \\ 
2436572.31340  &    1959-01-03  &    19:25  &    103 a-O  &    ...   &     SP   &     78   &         $B$  = 10.90 &   0.10   &       $B$ $>$15.86  &  0.10    &      $*$           &         \\ 
2436573.34808  &    1959-01-04  &    20:15  &    103 a-O  &    ...   &     SP   &     81   &         $B$  = 10.90 &   0.10   &       $B$ $>$17.15  &  0.10    &      $*$           &         \\
\hline
\end{tabular}
\end{tiny}
%\tablefoot{
%\tablefoottext{a}{A},\tablefoottext{b}{B},\tablefoottext{c}{C},\tablefoottext{d}{D}\\
%}
\end{table*}  
\setcounter{table}{0}
\begin{table*}
\caption{Continued.} 
\begin{tiny}
\begin{tabular}{l|cc|cccc|cc|cc|cc|}
\hline
\hline
  &   &   &   &   &  &  & \multicolumn{2}{c|}{NY Ori} & \multicolumn{2}{c|}{V1143 Ori} & \multicolumn{2}{c}{V1118 Ori} \\
HJD  &  date  &  UT  &  emulsion & filter  & tel & plate & mag & error & mag & error & mag & error \\
\hline                           
2437319.31944  &    1961-01-19  &    19:35  &    103 a-O  &    Liq   &     SP   &     1316 &         $B$  = 10.69 &   0.20   &       $ $           &          &      $*$           &         \\ 
2437319.39444  &    1961-01-19  &    21:23  &    103 a-O  &    Liq   &     SP   &     1318 &         $B$  = 10.69 &   0.20   &       $ $           &          &      $*$           &         \\ 
2437319.44582  &    1961-01-19  &    22:37  &    103 a-E  &    RG 1  &     SP   &     1319 &         $R$  = 10.15 &   0.10   &       $R$  = 15.61  &  0.10    &      $R$ $>$13.42  &  0.20   \\ 
2437336.32237  &    1961-02-05  &    19:41  &    103 a-E  &    RG 1  &     SP   &     1354 &         $R$  = 10.15 &   0.10   &       $R$  = 15.56  &  0.10    &      $R$ $>$15.13  &  0.10   \\ 
2437339.32144  &    1961-02-08  &    19:40  &    103 a-E  &    RG 1  &     SP   &     1372 &         $R$  = 10.15 &   0.10   &       $R$  = 15.56  &  0.10    &      $R$ $>$15.13  &  0.10   \\ 
2437342.30452  &    1961-02-11  &    19:16  &    103 a-E  &    RG 1  &     SP   &     1397 &         $R$  = 10.15 &   0.10   &       $R$  = 15.61  &  0.10    &      $R$ $>$15.13  &  0.10   \\
\hline
\end{tabular}
\end{tiny}
\end{table*}   
\setcounter{table}{0}
\begin{table*}
\caption{Continued.} 
\begin{tiny}
\begin{tabular}{l|cc|cccc|cc|cc|cc|}
\hline
\hline
  &   &   &   &   &  &  & \multicolumn{2}{c|}{NY Ori} & \multicolumn{2}{c|}{V1143 Ori} & \multicolumn{2}{c}{V1118 Ori} \\
HJD  &  date  &  UT  &  emulsion & filter  & tel & plate & mag & error & mag & error & mag & error \\
\hline  
2438769.39855  &    1965-01-08  &    21:28  &    103 a-O  &    ...   &     SP   &     5207 &         $V$  = 10.45 &   0.10   &       $V$ $>$15.94  &  0.10    &      $*$           &         \\ 
2438771.49705  &    1965-01-10  &    23:50  &    103 a-O  &    ...   &     SP   &     5214 &         $V$  = 10.45 &   0.10   &       $ $           &          &      $*$           &         \\                                                                                                                    
2438782.33389  &    1965-01-21  &    19:56  &    103 a-O  &    ...   &     SP   &     5227 &         $V$  = 10.45 &   0.10   &       $V$ $>$15.94  &  0.10    &      $V$ $>$15.81  &  0.10   \\ 
2438784.38236  &    1965-01-23  &    21:06  &    103 a-O  &    ...   &     SP   &     5235 &         $V$  = 10.45 &   0.10   &       $V$ $>$15.94  &  0.10    &      $V$ $>$15.81  &  0.10   \\ 
2438785.40660  &    1965-01-24  &    21:41  &    103 a-O  &    ...   &     SP   &     5251 &         $V$  = 10.45 &   0.10   &       $V$ $>$15.94  &  0.10    &      $V$ $>$15.81  &  0.10   \\ 
2438786.36000  &    1965-01-25  &    20:34  &    103 a-O  &    ...   &     SP   &     5260 &         $V$  = 10.45 &   0.10   &       $V$ $>$14.79  &  0.10    &      $V$ $>$15.81  &  0.10   \\
\hline  
\end{tabular}
\end{tiny}
\end{table*} 
\setcounter{table}{0}
\begin{table*}
\caption{Continued.} 
\begin{tiny}
\begin{tabular}{l|cc|cccc|cc|cc|cc|}
\hline
\hline
  &   &   &   &   &  &  & \multicolumn{2}{c|}{NY Ori} & \multicolumn{2}{c|}{V1143 Ori} & \multicolumn{2}{c}{V1118 Ori} \\
HJD  &  date  &  UT  &  emulsion & filter  & tel & plate & mag & error & mag & error & mag & error \\
\hline                                         
2439915.28576  &    1968-02-28  &    18:51  &    IN Sen   &    RG 5  &     SG   &     1447 &         $I$  = 9.56  &   0.10   &       $I$  = 14.11  &  0.10    &      $I$  = 14.51  &  0.10   \\ 
2439915.30450  &    1968-02-28  &    19:18  &    103 a-O  &    GG 13 &     SG   &     1448 &         $B$  = 10.83 &   0.10   &       $B$  = 17.66  &  0.10    &      $*$           &         \\                                                                             
2440157.66927  &    1968-10-28  &    03:58  &    103 a-E  &    RG 1  &     SG   &     1946 &         $R$  = 10.12 &   0.10   &       $R$  = 15.24  &  0.10    &      $R$  = 16.19  &  0.10   \\ 
2440185.61886  &    1968-11-25  &    02:44  &    103 a-O  &    ...   &     SG   &     1980 &         $B$  = 10.90 &   0.10   &       $B$  = 17.54  &  0.10    &      $*$           &         \\ 
2440203.46749  &    1968-12-12  &    23:06  &    103 a-O  &    ...   &     SG   &     2036 &         $B$  = 10.83 &   0.10   &       $B$  = 17.51  &  0.10    &      $*$           &         \\ 
2440208.57365  &    1968-12-17  &    25:39  &    103 a-O  &    ...   &     SG   &     2056 &         $B$  = 10.83 &   0.10   &       $B$ $>$17.46  &  0.10    &      $*$           &         \\
\hline  
\end{tabular}
\end{tiny}
\end{table*} 
\setcounter{table}{0}
\begin{table*}
\caption{Continued.} 
\begin{tiny}
\begin{tabular}{l|cc|cccc|cc|cc|cc|}
\hline
\hline
  &   &   &   &   &  &  & \multicolumn{2}{c|}{NY Ori} & \multicolumn{2}{c|}{V1143 Ori} & \multicolumn{2}{c}{V1118 Ori} \\
HJD  &  date  &  UT  &  emulsion & filter  & tel & plate & mag & error & mag & error & mag & error \\
\hline     
2441713.38255  &    1973-01-30  &    21:07  &    103 a-O  &    GG 13 &     SG   &     6185 &         $B$  = 10.83 &   0.10   &       $B$ $>$17.15  &  0.10    &      $*$           &         \\ 
2441713.39991  &    1973-01-30  &    21:32  &    IN Sen   &    RG 5  &     SG   &     6186 &         $I$  = 9.56  &   0.10   &       $ $           &          &      $ $           &         \\                                                                                                                 
2441980.61214  &    1973-10-25  &    02:36  &    IN Sen   &    RG 5  &     SG   &     6773 &         $I$  = 9.38  &   0.10   &       $I$  = 13.98  &  0.10    &      $I$  = 14.73  &  0.10   \\ 
2441984.47277  &    1973-10-28  &    23:15  &    103 a-E  &    RG 1  &     SG   &     6817 &         $R$  = 10.09 &   0.20   &       $R$  = 15.51  &  0.20    &      $R$ $>$15.13  &  0.20   \\ 
2442015.58626  &    1973-11-29  &    01:57  &    103 a-D  &    GG 14 &     SG   &     6887 &         $V$  = 10.42 &   0.10   &       $V$ $>$15.94  &  0.10    &      $V$ $>$16.19  &  0.10   \\ 
2442019.50851  &    1973-12-03  &    00:05  &    IN Sen   &    RG 5  &     SG   &     6915 &         $I$  = 9.74  &   0.10   &       $I$ $>$13.57  &  0.10    &      $I$ $>$14.22  &  0.10   \\
\hline   
\end{tabular}
\end{tiny}
\end{table*}
\setcounter{table}{0}
\begin{table*}
\caption{Continued.} 
\begin{tiny}
\begin{tabular}{l|cc|cccc|cc|cc|cc|}
\hline
\hline
  &   &   &   &   &  &  & \multicolumn{2}{c|}{NY Ori} & \multicolumn{2}{c|}{V1143 Ori} & \multicolumn{2}{c}{V1118 Ori} \\
HJD  &  date  &  UT  &  emulsion & filter  & tel & plate & mag & error & mag & error & mag & error \\
\hline    
2444607.48216  &    1981-01-02  &    23:28  &    103 a-E  &    RG 1  &     SG   &     10758 &        $R$  = 9.89  &   0.10   &       $R$  = 15.66  &  0.10    &      $R$  = 15.34  &  0.10   \\ 
2444609.39457  &    1981-01-04  &    21:22  &    103 a-O  &    GG 13 &     SG   &     10770 &        $B$  = 10.83 &   0.10   &       $B$ $>$17.15  &  0.10    &      $*$           &         \\                                                                                                                  
2444630.40444  &    1981-01-25  &    21:38  &    103 a-O  &    GG 13 &     SG   &     10818 &        $B$  = 10.83 &   0.10   &       $B$ $>$17.15  &  0.10    &      $*$           &         \\ 
2444630.42180  &    1981-01-25  &    22:03  &    103 a-D  &    GG 14 &     SG   &     10819 &        $V$  = 10.38 &   0.10   &       $V$  = 16.02  &  0.10    &      $V$ $>$16.19  &  0.10   \\ 
2444630.44263  &    1981-01-25  &    22:33  &    103 a-E  &    RG 1  &     SG   &     10820 &        $R$  = 9.89  &   0.10   &       $R$  = 15.66  &  0.10    &      $R$ $>$15.13  &  0.10   \\ 
2444630.46832  &    1981-01-25  &    23:10  &    IN Sen   &    RG 5  &     SG   &     10821 &        $I$  = 9.56  &   0.10   &       $I$  = 14.34  &  0.10    &      $I$  = 14.79  &  0.10   \\
\hline   
\end{tabular}
\end{tiny}
\end{table*}
\setcounter{table}{0}
\begin{table*}
\caption{Continued.} 
\begin{tiny}
\begin{tabular}{l|cc|cccc|cc|cc|cc|}
\hline
\hline
  &   &   &   &   &  &  & \multicolumn{2}{c|}{NY Ori} & \multicolumn{2}{c|}{V1143 Ori} & \multicolumn{2}{c}{V1118 Ori} \\
HJD  &  date  &  UT  &  emulsion & filter  & tel & plate & mag & error & mag & error & mag & error \\
\hline       
2448595.53490  &    1991-12-05  &    00:43  &    103 a-O  &    GG 13 &     SG   &     15099 &        $B$  = 10.83 &   0.10   &       $B$  = 17.18  &  0.10    &      $*$           &         \\ 
2448651.43058  &    1992-01-29  &    22:16  &    103 a-O  &    GG 13 &     SG   &     15153 &        $B$  = 10.83 &   0.10   &       $B$  = 17.46  &  0.10    &      $B$  = 17.16  &  0.10   \\                                                                                                              
2448651.45489  &    1992-01-29  &    22:51  &    IN Sen   &    RG 5  &     SG   &     15154 &        $I$  = 9.56  &   0.10   &       $I$  = 14.18  &  0.10    &      $I$  = 14.18  &  0.10   \\ 
2448682.35092  &    1992-02-29  &    20:25  &    IN Sen   &    RG 5  &     SG   &     15218 &        $I$  = 9.56  &   0.10   &       $I$  = 13.71  &  0.10    &      $I$  = 14.22  &  0.10   \\ 
2448686.32349  &    1992-03-04  &    19:46  &    IN Sen   &    RG 5  &     SG   &     15236 &        $I$  = 9.38  &   0.10   &       $ $           &          &      $ $           &         \\ 
2448974.47643  &    1992-12-17  &    23:19  &    103 a-O  &    GG 13 &     SG   &     15339 &        $B$  = 10.76 &   0.10   &       $B$  = 16.97  &  0.10    &      $*$           &         \\
\hline     
\end{tabular}
\end{tiny}
\end{table*}

\section{Analysis and discussion}

\subsection{V1118 Ori}

Examining its light curve in $R$ and $I$ bands, depicted in Figure~\ref{V1118_light:fig}, three major ($\Delta$m \gapprox~ 2 mag) flare-up events
are recognizable: a first one in 1966 February (R band), a second one in 1983 November ($R$, $I$ bands), and a third one in 1989 February ($B$, $R$, and $I$ bands).
This first event is mentioned here for the first time; however, as indicated in Table~\ref{photometry:tab} (HJD 2439172), it was a sudden flaring more than an EXor 
typical outburst, since the $R$ band brightness increased by 2.5 mag in 40 minutes, representing a very fast onset of a flare lasting less than one months (see $I$ band values in the same Table). Similar episodes, usually associated with very active chromospheres and hence detected at higher energies, are not commonly observed in either this source or 
other EXors, but, unfortunately, this episode has been sampled only in the $R$ band, thus preventing any colour analysis (see below).

However, it apparently marks the beginning of a long quiescence period from 1966 to 1981, as testified by the subsequent monitoring
in the $I$ band, whose temporal coverage suggests the absence of potential outbursts lasting more than 1 yr. Two observational gaps, 
slightly longer than 1 yr, indeed exist in 1973-74 and 1975-76 (see Table~\ref{photometry:tab}), so we cannot exclude that an outburst
might have occurred during one or both these observational gaps. Remarkably, this quiescence period (about 15 yrs) would be the longest known 
so far (Lorenzetti et al. 2015), thus its relevance is twofold: firstly, it poses a compelling constraint 
on the existence of a recurrent interval between two bursts; secondly, it allows us to define a reliable quiescent value of $I$ = 14.6 mag
(see Table~\ref{ranges:tab}). 
The second and third events are already mentioned in the literature by Parsamian et al. (1993), Garc\'{i}a Garc\'{i}a \& Parsamian (2000), who gave only the B and V magnitudes: our photometry is in agreement
with their values, but provides, for the first time, the $R$ and $I$ magnitudes, as well. In particular for the third event, by considering only the simultaneous plates 
(i.e. taken at a maximum temporal distance of one day), we are able to build a colour-magnitude plot $B$ vs.[$B-I$] practically unaffected by short time fluctuations 
and shown in the left panel of Figure~\ref{B_BI_plot:fig}. 
Still unpublished data (Giannini et al. in preparation), that we collected during
the very recent 2015-2016 outburst (Giannini et al. 2016) are shown on the figure, as are data from the current plate study. In both cases, the
source presents the very common behaviour to become bluer when brightening, but definitely not following the extinction law (depicted by the arrow).
These colours provide a further confirmation that the increase of brightness of V1118 Ori is accretion- more than extinction-driven (Lorenzetti et al. 2015).
Moreover, by considering the two distributions of open and solid data points, one could suspect that the different origin of the two data set (CCD and plates) has a 
role in determining the colour segregation, but this is not true as confirmed by the green data points relative to the 2005 outburst (Audard et al. 2005). 
The 2005 data perfectly overlap the plate photometry presented here, thus testifying on the one hand the reliability of the Asiago plates, on the second hand the fact that different bursts 
may have different colours and intensities.

\subsection{V1143 Ori} 

Two bursts are recognizable in our light curves (see Figure~\ref{V1143_light:fig} and also Table~\ref{photometry:tab}): the first appears as a steep rise up to a maximum in January 1983, 
followed by a slower declining; the second occurred in January 1993 and presents the same rapid increase (less than one month).
Mention of both events was already given by Parsamian \& Gasparian (1987) and Mampaso \& Parsamian (1995), respectively, but quiescence data
in the same bands are not provided, so preventing any comparative colour analysis. Our plate data, instead, allow this comparison, given in the 
right panel of Figure~\ref{B_BI_plot:fig}. This latter indicates the two bursts were practically identical in both $B$ amplitude and [$B-I$] colour, and,
more importantly, that this two bursts follow exactly the extinction curve with a variation corresponding to $\Delta$A$_V$ = 2 mag. This result, if confirmed
by future observations, could indicate V1143 Ori is not a genuine EXor, since both its historical outbursts are severely contaminated by
extinction effects. As for V1118, our data provide a good overall sampling of the quiescent phase, in fact the $I$-band photometry fills the temporal gap
1972-1977 uncovered in the $BVR$ bands. Our quiescence photometry agrees with the sparse values listed so far in the literature (Herbig, 2008 and references therein).  

\subsection{NY Ori}  

As previously mentioned, the measurements of NY Ori (Table~\ref{photometry:tab}) actually refer to the combination of the EXor source and the close star V566 Ori. 
Optical light curves (Figure~\ref{NY_light:fig}) seem dominated
by the brighter V566 Ori in the bands $BVR$, while in the $I$ band the EXor becomes to prevail. The shortest wavelength colours are compatible with an early type star (SpT A-F), whereas at longer wavelength colours become redder and are those typical of a late type source. A confirmation of that stems from the $JHK$  near-infrared
photometry carried out at our telescope at Campo Imperatore (D'Alessio et al. 2000). For V566 Ori, we obtained $J$ = 9.73, $H$ = 9.70, and $K$ = 9.70,
in agreement with 2MASS results ($J$ = 9.75, $H$ = 9.74, and $K$ = 9.73); while our IR monitoring of the EXor NY Ori provided evidence of both redder colours and some variability: $J$ (9.8 - 10.6), $H$ (8.9 - 9.4), and $K$ (8.1 - 8.5). At the same telescope, we obtained low-resolution ($\mathcal{R}$ $\sim$ 200) near-IR spectra
(1-2.5 $\mu$m) of both sources. V566 Ori is characterised by strong HI absorption lines (Paschen and Brakett series), while the EXor appears as a
typical emission-line object (Lorenzetti et al. 2009). Finally, we note in Figure~\ref{NY_light:fig} how the temporal coverage in $V$ band fills the 
gap of $I$ band observations in the period 1961-1967, so we can conclude that during our 40 yrs monitoring no major outburst occurred with a duration 
longer than six months. Such a conclusion is based on the analysis of the $I$ ligthcurve, namely that in which both stars have a comparable brightness.

%%%%%%%%%%%%%%   TABLE 2 - TIMESCALE %%%%%%%%%%%%%%%%%%%%%%%%%%%%%%%%%

\begin{table}
\caption{Ranges of photometric variability. For each source and each band, we list the number of observations (col.3), the median, which basically indicates the magnitude in quiescence (col.4), the standard deviation data point distribution (col.5), and the magnitude corresponding to the peak brightness (col.6). \label{ranges:tab}} 
\begin{center} 
\medskip
{
%\scriptsize
\begin{tabular}{lcc|ccc}
\hline
Source          &  Band     & N$_{obs}$ & Median & $\sigma$ &Peak  \\ 
                &           &           &          \multicolumn{3}{c}{(mag)    }     \\
\hline
\hline
V1118 Ori       &   $B$       &  13       &  16.62        &  0.25   &  14.16            \\
                &   $V$       &   2       &  16.02        &   ...   &   ...             \\ 
                &   $R$       &  33       &  15.51        &  0.64   &  13.33            \\  
                &   $I$       &  69       &  14.56        &  0.24   &  11.63            \\   
                \hline       
V1143 Ori       &   $B$       &  67       &  17.49        &  0.25   &  14.35            \\
                &   $V$       &  52       &  16.71        &  0.44   &  13.76            \\ 
                &   $R$       &  47       &  15.40        &  0.41   &  13.10            \\  
                &   $I$       &  81       &  14.18        &  0.15   &  13.31            \\  
                \hline
NY Ori          &   $B$       &  180      &  10.83        &  0.10   &  10.56            \\
                &   $V$       &  93       &  10.45        &  0.04   &  10.35            \\ 
                &   $R$       &  65       &  10.09        &  0.11   &  9.66             \\  
                &   $I$       &  102      &  9.56         &  0.14   &  9.20             \\  
\hline
\end{tabular}}
\end{center}

%\medskip
%$^a$ This sampling was obtained also on hours time-scale.
%\medskip 
\end{table}    
%%%%%%%%%%%%%%%%%%%%%%%%%%%%%%%%%%%%%%%%%%%%%%%%%%%%%%%%%%%%%%%%%%%%%%%

\section{Concluding remarks}

Archival plate analysis is a powerful tool to investigate the historical behaviour of EXor stars, and to infer on their nature.

We investigated the Asiago Schmidt plate collection for observations of the Orion Nebula Cluster where the three EXors sources V1118 Ori, V1143 Ori, and NY Ori are located.
Observations of this region were repeatedly carried out at Asiago over a timespan of about 40 years since 1958 and the three sources were acquired on the same plates.

We provide one of the best-sampled photometric dataset ever obtained of the quiescent phase of the three targets. V1118 and V1143 present, in quiescence,
a level of modest variability (0.2-0.4 mag) that is comparable with that of classical T Tauri stars.

For V1118 Ori, two already known outbursts are detected together with a newly discovered flare-up of 2.5 mag ($R$ band), which brightened up in only 40 min and lasted
less than one month. Our data most likely ascertain the longest quiescence period known so far (about 15 yrs): this result on the one hand does not support the existence of a recurrent 
period between the outbursts, on the other hand allows us to define a reliable value of the quiescence brightness. The colour analysis confirms the 
robustness of the plate photometry and rules out the extinction as the main origin of the brightness variations.

For V1143 Ori, two outbursts already mentioned in the past literature are found, but the complete set of colours are given here for the first time.
This is relevant, since the colour analysis demonstrates as both outbursts, at variance with those of V1118, can be accounted for with pure 
extinction variations, putting some doubts on the EXor nature of V1143.

As for NY Ori, this target and the close source V566 Ori cannot be resolved in our plates, so that the given photometry refers to the combined pair.
Nevertheless, as the two objects have a similar brightness in the I band, we can conclude that the EXor source did not undergo any major outburst ($>$ 2 mag) during the 40 yr monitoring.

\begin{acknowledgements}
R.J.S. thanks the INAF Astronomical Observatory of Padova for the 
hospitality during the stay in Asiago and for permission to use the 
historical plate archive of the Asiago Observatory. This work was 
supported in part by the Croatian Science Foundation under the project 
6212 Solar and Stellar Variability and by the University of Rijeka 
under the project number 13.12.1.3.03.
\end{acknowledgements}

\end{document}